\documentclass[showpacs,preprintnumbers,twocolumn,superscriptaddress]{revtex4}
\usepackage{amssymb}
\usepackage[centertags]{amsmath}
\usepackage[paperwidth=210mm,paperheight=297mm,centering,hmargin=1.5 cm,vmargin=2.3 cm]{geometry}
\usepackage{txfonts}
\usepackage{epsfig}
\usepackage{bm}
\usepackage{graphicx,graphics}
\usepackage{float}
\begin{document}

\title{ Probing charge correlations of quark gluon plasma by identified two-hadron rapidity correlations in ultra-relativistic AA collisions }

\author{Jun Song}
\affiliation{Department of Physics, Jining University, Shandong 273155, China}

\author{Feng-lan Shao}
\email{shaofl@mail.sdu.edu.cn}  \affiliation{Department of Physics, Qufu Normal University, Shandong 273165, China}

\begin{abstract}
  We propose a new kind of two-particle correlation of identified hadrons in longitudinal rapidity space, called $G_{\alpha\beta}(y_{\alpha},y_{\beta})$, which can reflect clearly the charge correlations of hot quark system produced in AA collisions at LHC energies.
  It is derived from the basic scenario of quark combination mechanism of hadron production. Like the elliptic flow of identified hadrons at intermediate transverse momentum, this correlation is independent of the absolute hadronic yields but dependent only on the flavor compositions of hadrons, and thus exhibits interesting properties for different kinds of hadron species.  We suggest the measurement of this observable in AA collisions at LHC to gain more insights into the charge correlation properties of produced hot quark matter.
\end{abstract}
\pacs{25.75.Gz,25.75.-q}
\maketitle

Correlations between different identified hadrons in momentum space were sensitive probes of prompt hadron production dynamics in different high energy reactions already in 1980s - 1990s \cite{Andersson1983,webber1984,drifard1980,Aihara1984,Aihara1986,Acton1993,szg1997}.
The experimental data of baryon-baryon and baryon-antibaryon correlations in rapidity space in $e^{+}e^{-}$ annihilations \cite{Aihara1984,Aihara1986,Acton1993} provided important test of existing phenomenological models of non-perturbative hadronization process.
In ultra-relativistic heavy ion collisions, correlation measurements and studies continue to serve as an indispensable means in exploring the properties of strong-interacting quark gluon plasma (QGP) produced in collisions \cite{starBtB03,hirano03,starHBT,henze96HBT,pratt09hbt,Bass_clockHd,SJeon02,Bialas04,ChengS04,Bozek05StatResonance}.
Recent studies of low $p_T$ domain correlations in both experimental and theoretical sides \cite{Bass_clockHd,SJeon02,Bialas04,ChengS04,Bozek05StatResonance,LNbf09,Fjh11bf,SJ12FB,StarBF130,NA49SCden,NA49REden,StarBF200AApp,StarBF200Scaling,alice13bf} are mainly of charged particles in terms of balance functions while those of two-particle correlations of identified hadrons are relatively less concerned, and are mainly of charged pion and kaon which suffer greatly from decay contamination.
The latest progress is the thermal analysis of Scott Pratt \cite{Pratt11_bf}.
With the rapid development of high precision LHC experiments and deep running of RHIC, the underlying physics of two-hadron correlations in the context of AA collisions is urgently needed to interpret, hoped as effectively as hadronic  elliptic flow, what can we learn from identified hadron correlations, especially what QGP properties can be extracted from correlations of which particular identified hadrons.

In this paper, we propose the following two-hadron correlation function in longitudinal rapidity space as a new observable in AA collisions at LHC energies,
\begin{equation}
G_{\alpha\beta}(y_{1},y_{2})  = \frac{  \Big\langle [n_{\alpha}(y_1)- n_{\bar{\alpha}}(y_1)][ n_{\beta}(y_2)- n_{\bar{\beta}}(y_2)] \Big\rangle }
 {\langle n_{\alpha}(y_1)\rangle \langle n_{\beta}(y_2)\rangle},
 \label{Gab}
\end{equation}
which measures directly the correlation between two hadronic species by a symmetrical combination of $\alpha \beta$, $\bar{\alpha} \bar{\beta}$,
${\alpha} \bar{\beta}$ and $\bar{\alpha} {\beta}$ correlations but can reflect in a quite clear manner the properties of conserved charge correlations of the quark system produced in collisions.
Here, angle brackets denote event average and $n_\alpha(y_1)$ the number of hadron $\alpha$  at $y_1$. We confine ourself to the situation of zero baryon number density, e.g. the central plateau region of collisions as a good approximation, otherwise one would subtract terms $ \langle n_{\alpha}(y_1)- n_{\bar{\alpha}}(y_1)\rangle
 \langle n_{\beta}(y_2)- n_{\bar{\beta}}(y_2) \rangle$ in the numerator.
In the denominator of Eq. (\ref{Gab}), we put the product of two hadron yields rather than the either one, which is the major difference from previous balance functions \cite{drifard1980,Bass_clockHd,Pratt11_bf}.
At first sight, this correlation is hadronic yield dependent since it seems to be in the magnitude of $1/n_{\alpha}$ or $1/n_{\beta}$.
In fact, however, it is not the case but presents a clean quark level insight as we interpret Eq. (\ref{Gab}) in the framework of the quark combination mechanism (QCM) of hadron production.

QCM describes the production of hadrons as quark system hadronizes by the combination of neighboring quarks and antiquarks in phase space.
Reducing QCM to one-dimensional longitudinal rapidity space and considering that the averaged rapidity interval between neighboring quarks at LHC energies is the order of $10^{-3}$, i.e. thousand quarks per unit rapidity, the combination of neighboring quarks into a hadron can be treated as an ideal equal $y$ combination to make the study illustrative and insightful and meanwhile keep the good numerical approximation.
Then, we have the single hadron rapidity distribution, e.g. for the produced meson $M_i(q_1\bar{q}_2)$
\begin{equation}
  \mathcal{F}_{M_i}(y) = \mathcal{P}_{q_1\bar{q_2},M_i}(y)\ \  f_{q_1\bar{q}_2}(y),  \label{myield}
\end{equation}
and similar for baryonic $\mathcal{F}_{B}(y)$.
Here, $f_{q_1\bar{q}_2}(y)$ is the density of $q_1\bar{q}_2$ pairs at rapidity $y$ in system just before hadronization.
$\mathcal{P}_{q_1\bar{q_2},M_i}$ denotes the probability of a ${q_1\bar{q}_2}$ pair combining into a meson.
We also consider the two-hadron joint distributions $\mathcal{F}_{\alpha \beta}(y_1,y_2)$ since $G_{\alpha \beta}$ essentially measures $(\mathcal{F}_{\alpha \beta}+\mathcal{F}_{\bar{\alpha} \bar{\beta}}-\mathcal{F}_{\bar{\alpha} \beta}-\mathcal{F}_{\alpha \bar{\beta}})/ \mathcal{F}_{\alpha}\mathcal{F}_{\beta}$, and have
\begin{equation}
\mathcal{F}_{M_i M_j}(y_1,y_2) = \mathcal{P}_{(q_1\bar{q_2})(q_3\bar{q_4}),M_i M_j}(y_1,y_2)\ \  f_{(q_1\bar{q_2})(q_3\bar{q_4})}(y_1,y_2)
  \label{byield}
\end{equation}
and similar for $\mathcal{F}_{MB}$ and $\mathcal{F}_{BB}$. $f_{(q_1\bar{q_2})(q_3\bar{q_4})}(y_1,y_2)$ are joint distribution of $q_1\bar{q_2}$ pair at $y_1$ and $q_3\bar{q_4}$ pair at $y_2$. $\mathcal{P}_{(q_1\bar{q_2})(q_3\bar{q_4}),M_i M_j}(y_1,y_2)$ is the joint production probability of $M_i M_j$ pair given a $q_1\bar{q_2}$ pair at $y_1$ and a $q_3\bar{q_4}$ pair at $y_2$, which has the factorization form $\mathcal{P}_{q_1\bar{q_2},M_i}(y_1) \, \mathcal{P}_{q_3\bar{q_4},M_j}(y_2)$ because of the locality of hadronization.
In the system without net charges we expect $\mathcal{P}_{q_1\bar{q_2},M_i} = \mathcal{P}_{\bar{q_1}q_2,\bar{M}_i}$ and $\mathcal{P}_{q_1q_2q_3,B_j} = \mathcal{P}_{\bar{q}_1\bar{q}_2\bar{q}_3,\bar{B}_j}$ in charge conjugation symmetry.

Kernel $\mathcal{P}$ involves the complex non-perturbative hadronization dynamics and are little known presently from first principles.
Fortunately, by adopting Eq. (\ref{Gab}) we don't need to known the precise form of $\mathcal{P}$, thereby avoiding those still debated issues or puzzles caused by unclear hadronization dynamics, e.g. energy and entropy conversation in low $p_T$ domain \cite{ckm03,nonaka05,hwa04,biro07entropy,sj10s},  to make the general conclusion.
Using Eqs. (\ref{myield}-\ref{byield}) we obtain that $G_{\alpha \beta}$ actually measures the $(f^{(q)}_{\alpha\beta}+f^{(q)}_{\bar{\alpha} \bar{\beta}}-f^{(q)}_{\bar{\alpha} \beta}-f^{(q)}_{\alpha\bar{\beta}})/f^{(q)}_{\alpha}f^{(q)}_{\beta}$ where $f^{(q)}_{\alpha}$ and $f^{(q)}_{\alpha\beta}$ denote the corresponding multi-quark distributions for single $\alpha$ and joint $\alpha\beta$ productions in Eqs. (\ref{myield}-\ref{byield}), respectively. Then we immediately get
\begin{equation}
  G_{\alpha\beta}(y_{1},y_{2}) = \frac{  \Big\langle [n^{(q)}_{\alpha}(y_1)- n^{(q)}_{\bar{\alpha}}(y_1)][ n^{(q)}_{\beta}(y_2)- n^{(q)}_{\bar{\beta}}(y_2)] \Big\rangle }
 {\langle n^{(q)}_{\alpha}(y_1)\rangle \langle n^{(q)}_{\beta}(y_2)\rangle} .
\end{equation}
 $n^{(q)}_{\alpha}(y)$ is the number of multi-quark pairs in a event, corresponding to theoretical $f^{(q)}_{\alpha}(y)$.
In meson $\alpha(q_1\bar{q}_2)$ case, $n^{(q)}_{\alpha}(y)=n_{q_1 \bar{q_2}}(y)=n_{q_1 }(y)n_{\bar{q_2}}(y)$ is the number of $q_1 \bar{q_2}$ pairs at rapidity $y$. In baryon $\alpha(q_1 q_2 q_3)$ case, $n^{(q)}_{\alpha}(y)=n_{q_1 q_2 q_3}(y)$ is the number of $q_1 q_2 q_3$ combinations which satisfies $ n_{q_1 q_2 q_3}(y)\approx n_{q_1}(y)n_{q_2}(y)n_{q_3}(y)$ in large quark number limit. $n_q(y)$ is the number of $q$ flavor quarks at $y$. To second order in the fluctuations of quark numbers, we have
\begin{equation}
 G_{\alpha\beta}(y_{1},y_{2})=
  \sum_{f_1,f_2} A_{f_1 f_2} \  \frac{C_{f_1f_2}(y_1,y_2)}{\langle n_{f_1}(y_1)\rangle \langle n_{f_2}(y_2)\rangle }
   \label{tgab}
\end{equation}
where $A_{f_1 f_2}=(n_{\alpha,f_1}-n_{\bar{\alpha},f_1})(n_{\beta,f_2}-n_{\bar{\beta},f_2})$.
Here, $f_1$ and $f_2$ run over all kinds of quark flavors. $n_{\alpha,f_1}$ is the number of quark $f_1$ contained in hadron $\alpha$.
$C_{f_1 f_2}(y_1,y_2)=\Big\langle  n_{f_1}(y_1) n_{f_2}(y_2) \Big\rangle - \Big\langle n_{f_1}(y_1) \Big\rangle \Big\langle n_{f_2}(y_2)  \Big\rangle$ is the ordinary two-point correlation function.
Clearly, $G_{\alpha\beta}(y_{1},y_{2})$ depends only on the quark level correlations as well as the flavor compositions of hadrons $\alpha$ and $\beta$, independent of the absolute yields of two hadrons.
We emphasis that this result is independent of the precise form of kernel $\mathcal{P}$ and, therefore, is a general result relating only to the basic scenario/dynamics of QCM.

As apply the approximation of zero baryon number density to the central plateau region of AA collision at LHC energies, we further assume a Bjorken longitudinally boost-invariance \cite{bjorken1983} for the system, and then the two point correlation functions depend only on $\Delta y  = y_2 - y_1$  rather than on $y_1$ and $y_2$ individually. Since the rapidity density of particles is uniform in this case, we use $n_\alpha$ and $n_f$ to denote the rapidity density of hadron $\alpha$ and quark $f$, respectively.

One of the advantages of $G_{\alpha\beta}$ is that it can conveniently relate to the correlation of conserved charges in quark system.
Here we consider the system made up mainly of the three quark flavors, i.e. up ($u$), down ($d$) and strange ($s$) quarks. There are three conserved charges in system, i.e.  baryon number ($B$), electric charge ($C$) and strangeness ($S$). Following the Ref \cite{Pratt11_bf}, we alternatively use the net number of up ($n_u-n_{\bar{u}}$), down ($n_d-n_{\bar{d}}$) and strange quarks ($n_s-n_{\bar{s}}$) instead of $B$, $C$ and $S$ charges because quark numbers are more convenient in hadronic correlations. Furthermore, charge correlations of quark system $G_{ab}(\Delta y)$ can be also described by  Eq.(\ref{Gab}). Under the isospin and charge conjugation symmetry, we have four independent correlation functions, i.e.,

\begin{eqnarray}
G_{uu}(\Delta y) &=& 2 \Big[ C_{uu}(\Delta y) -C_{u\bar{u}}(\Delta y) \Big] / \langle n_u\rangle ^{2}, \nonumber \\
G_{ss}(\Delta y) &=& 2 \Big[ C_{ss}(\Delta y) -C_{s\bar{s}}(\Delta y) \Big] / \langle n_s\rangle ^{2}, \label{gab_cqs} \\
G_{ud}(\Delta y) &=& 2 \Big[ C_{ud}(\Delta y) -C_{u\bar{d}}(\Delta y) \Big] / \langle n_u\rangle ^{2} , \nonumber \\
G_{us}(\Delta y) &=& 2 \Big[ C_{us}(\Delta y) -C_{u\bar{s}}(\Delta y) \Big] / \langle n_u\rangle \langle n_s \rangle \nonumber
\end{eqnarray}
for the quark system just before hadronization.

Substituting Eq.(\ref{gab_cqs}) into Eq.(\ref{tgab}), we finally get
\begin{equation}
  G_{\alpha\beta}(\Delta y)  = \sum_{f_1,f_2 =u,d,s} Q_{\alpha,f_1}Q_{\beta,f_2} G_{f_1 f_2}(\Delta y)
\end{equation}
where $Q_{\alpha,f_1}$ denotes the net number of $f_1$ in hadron $\alpha$.
The equation shows a direct and simple connection between the hadronic correlation and the charge correlation of quark system before hadronization.
There is no such concision if one adopt $C_{\alpha \beta}$ and existing balance functions.
In Table \ref{tab1}, we show $G_{\alpha \beta} (\Delta y) $ of various identified hadrons in terms of the charge correlations $G_{ab}(\Delta y)$ of the quark system.
From Eq. (\ref{Gab}) we have $G_{\alpha \beta}= G_{\beta\alpha }$, $G_{\alpha \bar{\beta}}= - G_{\alpha \beta}$ and $G_{\alpha \beta}= G_{\bar{\alpha} \bar{\beta}}$. Therefore, Table \ref{tab1} covers the correlations of major stable hadrons measured experimentally.

\begin{table*}
\caption{$G_{\alpha \beta}  (\Delta y)$ of directly-produced hadrons after hadronization in terms of $G_{ab} (\Delta y)$ of the quark system before hadronization. }
\renewcommand{\arraystretch}{1.2}
\centering
\begin{tabular}{c|cccccc} \hline \hline
$G_{\alpha \beta}$   &  $\pi^{+}$    & $p$       & $K^{+}$    & $\Lambda$        & $\Xi^{0}$        &   $\Omega^{-}$  \\  \hline
$\pi^{-}$            & $-2 G_{uu} +2 G_{ud}$   & & & & & \\
$\bar{p}$            & $- G_{uu} + G_{ud} $   & $-5 G_{uu} -4 G_{ud}$  & & & {\it transpose symmetric} & \\
$K^{-}$              & $- G_{uu} + G_{ud} $   & $-2 G_{uu} - G_{ud} + 3 G_{us}$  & $- G_{uu} + 2 G_{us} - G_{ss}$  & & & \\
$\bar{\Lambda}$     & 0                      & $-3 G_{uu} - 3 G_{ud} - 3 G_{us}$   & $- G_{uu} -  G_{ud} + G_{us} + G_{ss}$
                    & $- 2 G_{uu} - 2 G_{ud} - 4 G_{us} - G_{ss}$        &&     \\
$\bar{\Xi^{0}}$      & $- G_{uu} + G_{ud}$    &$- 2 G_{uu} - G_{ud} - 6 G_{us}$   &$- G_{uu} - G_{us} + 2 G_{ss}$
                     & $- G_{uu} - G_{ud} - 5 G_{us} - 2 G_{ss}$  & $- G_{uu}  - 4 G_{us} - 4 G_{ss}$   & \\
$\bar{\Omega^{+}}$   & 0                      & $-9 G_{us} $      & $- 3 G_{us} + 3 G_{ss} $   & $- 6 G_{us} - 3 G_{ss} $
                     &$- 3 G_{us} - 6 G_{ss} $  & $-9 G_{ss}$  \\
\hline \hline
\end{tabular}\label{tab1}
\end{table*}

There are several particularly interesting results in Table \ref{tab1}. First, we see that signs before $G_{ab}$ in $G_{MM}$ and $G_{MB}$ has both the positive and negative parts while in $G_{B\bar{B}}$ their signs are the same.
It arises from the flavor composition nature of $M(q\bar{q})$ and $B(qqq)$.
Second, we find  $G_{\pi^{+}\bar{\Lambda}}(\Delta y) =G_{\pi^{+}\bar{\Omega}}(\Delta y) =0$ which means their productions are independent of each other.  This is because $\pi^{+}$ is composed of $u$ and $\bar{d}$ and their correlations with quarks in ${\Lambda}$ and $\Omega$ are canceled out under the isospin symmetry. Third, we see $G_{\pi^{+}\bar{p}}(\Delta y) =G_{\pi^{+}K^{-}}(\Delta y)=G_{\pi^{+}\bar{\Xi}^0}(\Delta y) =-G_{uu}(\Delta y) + G_{ud}(\Delta y)$ in which strangeness correlations disappear. This is also because the correlation between $u$ in $\pi^{+}$ and strange quarks in $K^{-}, \bar{\Xi}^0$ cancels that part of $\bar{d}$ in pion and thus only light flavor correlations are left.  These results are independent of quark's $G_{ab}$ and thus can be regarded as the characteristic properties of QCM.

Besides, quark combination also predicts that $K^{+}$ is associated in production with  $\Lambda$, $\Xi^{0}$ and $\Omega^-$ rather than their anti-particles for the regular quark $G_{ab}$ such as those discussed later. This is because $+G_{ss}$ item always overwhelms numerically other parts in their correlation decompositions, see Table \ref{tab1}. It is a natural result since $K^{+}$ production consumes a $\bar{s}$ while remaining $s$ enters into a hyperon, thereby passing the $s\bar{s}$ quark correlation to these strange hadrons.

The fact of $G_{\alpha \beta}$ being the linear combination of a few $G_{ab}$, as shown in Table I,  also suggests that not only the correlation width measured in the past but also the correlation magnitude should be regarded as the significant and meaningful observations at LHC experiments.
On the other hand, these simple relations provide the possibility of extracting the charge correlation properties of quark system from the measurable correlations of identified hadrons.

Subsequently, we study the qualitative properties of $G_{ab}(\Delta y)$ of quark system and their identification by hadronic $G_{\alpha \beta}$ measurements. $G_{ab}$ contains two main ingredients: (a) intrinsic correlation between charge $a$ and charge $b$. Here we concentrate on the short range correlation (SRC) generated by interactions of thermal/soft partons during the later stage of QGP evolution.
(b) the global charge conservation (GCC) imposed on charge $a$ and $b$, respectively. It is mainly induced by the charge separation during the first fm/c of the collisions, and should has a long range characteristic in rapidity.
We take a Gaussian parametrization for their shapes in rapidity and have
\begin{equation}
  G_{ab}(\Delta y) = \frac{\chi_{ab}}{\langle n_a \rangle \langle n_b \rangle} \Bigg(  \frac{e^{- \frac{ {\Delta y } ^2}{2\sigma_{g}^2}}}{\sqrt{2\pi}\sigma_{g}} - \frac{e^{- \frac{ {\Delta y } ^2}{2\sigma_{s}^2}}}{\sqrt{2\pi}\sigma_{s}} \Bigg) .
\end{equation}
Here, $\sigma_{g}$ and $\sigma_{s}$ are GCC and SRC widths, respectively.
$\chi_{ab} \equiv \sum_{\alpha} q_{\alpha,a}q_{\alpha,b} \langle n_{\alpha} \rangle$ is the charge correlation matrix of the system \cite{Pratt11_bf} obtained by another decomposition of the numerator of Eq. (\ref{Gab}), which is used here to characterize/constrain the magnitude of $G_{ab}$. The summation $\alpha$ runs over all ingredient particles of the system, and $q_{\alpha,a}$ is the charge of type $a$ on $\alpha$.

Here, we consider three different charge correlation scenarios that are possible for the quark system at hadronization:
(1) the system consists of only quasi-free individual constituent quarks and antiquarks. $\chi_{ab}$ matrix is diagonal;
(2) among quarks and antiquarks in system there exist some tight correlation between quarks and antiquarks with different flavors in rapidity space. $\chi_{ab}$ has the negative off-diagonal matrix elements which will generally increase  $G_{MM}$ and $G_{BM}$ but decrease $G_{BB}$ in magnitudes;
(3) different from the former, there exist the tight correlation between two (anti-)quarks. $\chi_{ab}$ has the positive off-diagonal matrix elements which generally increase $G_{BB}$ but decrease $G_{MM}$ and $G_{BM}$.
Three different $G_{ab}$ scenarios can be identified by the measurement of the hadronic $G_{\alpha \beta}$. In Fig.1, we show calculations of $G_{p \bar{\Lambda}}(\Delta y)$, $G_{p \bar{\Xi}^0}(\Delta y)$ and $G_{p \bar{\Omega}^+}(\Delta y)$ at three scenarios as their  effective discrimination.

\begin{figure*}
  \includegraphics[width=\linewidth]{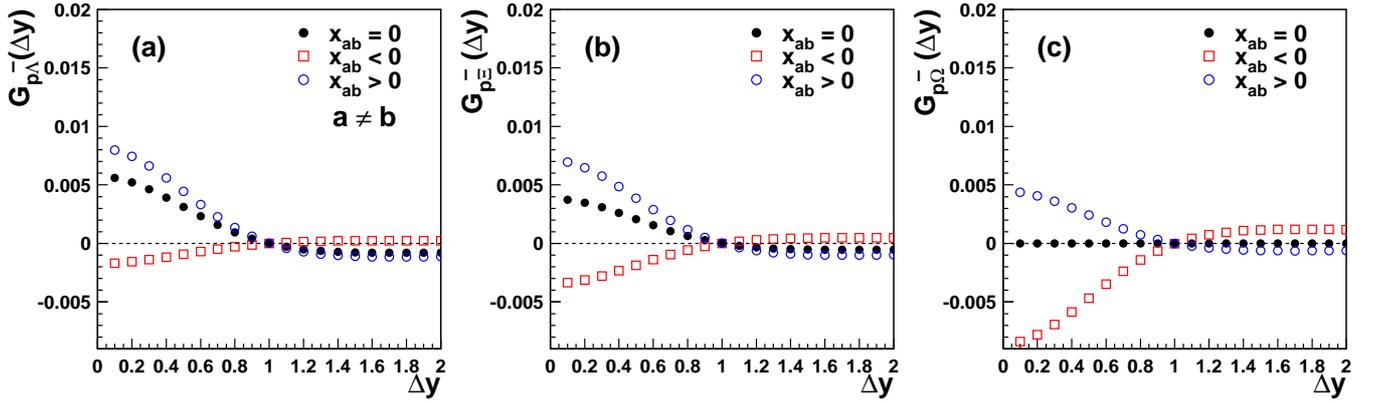}\\
  \caption{(Color online) $G_{p \bar{\Lambda}}(\Delta y)$ , $G_{p \bar{\Xi}^0}(\Delta y)$  and $G_{p \bar{\Omega}^+}(\Delta y)$ as the function of $\Delta y$, as $\chi_{ab}$ matrix is diagonal (filled circles), has negative off-diagonal elements (open squares) and has positive  off-diagonal elements (open circles).}\label{fig1}
\end{figure*}

In calculations, the quark rapidity densities are taken to be $\langle n_u \rangle = \langle n_d \rangle =710 $ and $\langle n_s \rangle = 290$, respectively, by using a specific combination model developed by Shandong group \cite{sdqcm} to fit the experimental data of rapidity density of pion and kaon in central Pb + Pb collisions at $\sqrt{s_{NN}} =2.76$ TeV \cite{alice13PiKp}.
For widths $\sigma_g$ and $\sigma_s$ we temporarily assume a flavor blind value for the purpose of qualitative analysis only.
GCC distribution width $\sigma_g$ is taken to be 3.8, fixed roughly by the data of pseudo-rapidity distribution of charged particles \cite{alice13dndeta}.
SRC width $\sigma_s$ is fixed to be 0.45 by the data of charge balance function \cite{alice13bf} in the collisions.
Matrix elements of diagonal $\chi_{ab}$ in scenario (1) is $\chi_{uu}=2 \langle n_u \rangle$ and $\chi_{ss} = 2 \langle n_s \rangle$.
For the scenario (2), we consider the case proposed in Ref \cite{Pratt11_bf}, $\chi_{us}\approx - \chi_{ss}/2$ and $\chi_{ud}\approx -\chi_{uu} + \chi_{ss}/2$ as an illustration of strong quark-antiquark correlations with different flavors in rapidity.
For the scenario (3), we take a thermal ansatz. The number of correlated two quarks in rapidity is assumed to be thermal distributed. The mass is the sum of quark masses. Fixing the hadronization temperature $T=165$ MeV and mass of individual quarks $m_u = 330$ MeV and $m_s = 500$ MeV, we estimate the magnitudes of off-diagonal elements to be $\chi_{ud} \approx 0.17 \chi_{uu}$ and $\chi_{us} \approx 0.26 \chi_{ss}$, respectively.

It can be seen from Fig.1 that in diagonal case of $\chi_{ab}$ the production between $p$ and $\bar{\Lambda}$, $\bar{\Xi}^0$ is associated. This is quite natural due to the local neutrality of net-$u$ charge in the system or in other words the light quark and light anti-quark production association. The production of $p$ and that of $\Omega$ are independent of each other in this case because of the vanishing $G_{us}$ component.
The positive off-diagonal $\chi_{ab}$ elements increase their production association magnitudes, even for the $p$ and $\bar{\Omega}^+$.
On the contrary, the large negative values of the off-diagonal $\chi_{ab}$ elements may change the sign of the $G_{p \bar{\Lambda}}$, $G_{p \bar{\Xi}^0}$ and even $G_{p \bar{\Omega}^+}$, which means the production between $p$ and these anti-hyperons is no longer be concomitant but repulsive.
We can see that $G_{p \bar{\Lambda}}(\Delta y)$ and $G_{p \bar{\Xi}^0}(\Delta y)$ can effectively discriminate the scenario (2) from others while $G_{p \bar{\Omega}^+}(\Delta y)$ is all powerful in three cases.

Decays of short-lived resonances are significant and complex contamination source on the study of charge correlation of QGP produced in collisions via the measurement of hadron correlations.
Here, we consider decays of the flavor SU(3) ground state hadrons.
Our calculations shows: (1) weak decays of hadrons contaminate to a large extent and even overwhelm the correlations of directly-produced hadrons in final observation, in particular for the correlation between pion and other hadrons $G_{\pi h}$.
(2) effects of strong and electromagnetic decays are varied with hadron species, but they are generally smaller than those of weak decays.
$G_{\pi h}$ correlations are still strongly influenced due to the still many decay channels into $\pi$.
$G_{K^+ K^-} $ correlation is influenced also by the magnitude decrease about $25\%$.
$G_{p K^-}$ correlation is slightly decreased less $10\%$.
$G_{K^+ \bar{\Lambda}}$, $G_{K^+ \bar{\Xi}^0}$, $G_{K^+ \bar{\Omega}^+}$, $G_{p \bar{\Lambda}}$, $G_{p \bar{\Xi}^0}$, $G_{p \bar{\Omega}^+}$, $G_{\Lambda \bar{\Lambda}}$, $G_{\Lambda \bar{\Xi}^0}$ and $G_{\Lambda \bar{\Omega}^+}$ are almost unchanged. Since the weak decays of strange hadrons can be corrected in experiments at LHC, these unchanged correlations are good observables.

There are two available methods measuring the hadron $G_{\alpha\beta}(\Delta y)$ in experiments.
The first is that adopted in $e^{+}e^{-}$ and $p\bar{p}$ reactions in the early years, i.e. choose hadrons $\alpha$ and $\bar{\alpha}$ at a specific rapidity, e.g. at $y=0$, as the test particles and then record rapidity distances between every hadron $\beta$ ($\bar{\beta}$) and test particles event-by-event.
The second is that used recently in balance function measurements in AA collisions. Considering the detectors have a finite acceptance rapidity window $y_w$, statistics of all hadrons $\alpha$, $\bar{\alpha}$, $\beta$ and $\bar{\beta}$ in this window generates the partial correlation function $G_{\alpha\beta}(\Delta y|y_w)$, and then divide it by the scale factor $1-\Delta y / y_w $ proposed in Ref \cite{SJeon02} to remove the finite window effects and restore the theoretical definition.

In summary, we have studied in QCM the two-particle correlation of identified hadrons in longitudinal rapidity space in ultra-relativistic AA collisions. We presented a new correlation function which can reflects more clearly than the balance functions the charge correlations of quark system produced in collisions. It is also useful to guide the future experimental measurements of charge correlations with focused goal at LHC.

Another point needed to address at last is gluon effects.
Although gluon does not carry conserved charges $B$, $C$ and $S$, it has definite nontrivial effects on hadronic correlations.
In QCM, gluon at hadronization is usually replaced by a pair of quark and antiquark, which will increase $\chi_{uu}$ and $\chi_{ss}$ of the system and therefore influence mostly the correlations between particle pairs with opposite quantum numbers.
We have considered implicitly the gluon effects in this paper since we directly start from the system of quarks and antiquarks.
We will study separately the effects of gluons at hadronization on hadronic correlation as well as fluctuations in future work.

{\it Acknowledgments.} The authors thank Z. T. Liang and R. Q. Wang for helpful discussions. The work is supported in part by the National Natural Science Foundation of China under grant 11175104, 11247202 and 11305076, and by the Natural Science Foundation of Shandong Province, China under grant ZR2011AM006 and ZR2012AM001.

\end{document}